\documentclass[aps,prd,twocolumn,superscriptaddress,showpacs,floatfix,handout,10pt]{revtex4-1}

\usepackage{CJKutf8} % 或者使用 \usepackage{CJK} 并在文档中使用 \begin{CJK*}{UTF8}{gbsn}...\end{CJK*} 环境
\usepackage{amsmath,amssymb,mathtools}
\usepackage{color}
\usepackage{graphicx}
\usepackage{subfigure}
\usepackage[colorlinks=true,linkcolor=blue, citecolor=red, urlcolor=blue, bookmarks]{hyperref}
\usepackage{multirow,makecell} %% multirow and multicolumn
\usepackage{textcomp}
\usepackage{wasysym}
\usepackage{ulem}

\def \be {\begin{equation}}
\def \ee {\end{equation}}
\def \ba {\begin{array}}
\def \ea {\end{array}}
\def \bea{\begin{eqnarray}}
\def \eea{\end{eqnarray}}
\def \nn {\nonumber}

\def \a {\alpha}
\def \b {\beta}

\def \G {\Gamma}
\def \d {\delta}
\def \D {\Delta}
\def \e {\epsilon}

\def \s {\sigma}

\def \r {\rho}

\def \cA {\mathcal A}

\def \cX {\mathcal X}
\def \cY {\mathcal Y}

\def \mX {\mathcal X}

\def \p {\partial}

\def \f {\frac}

\def \sr {\sqrt}

\def \inf {\infty}

\def \lag {\langle}
\def \rag {\rangle}

\def \ep {\mathrm{e}}
\def \ii {\mathrm{i}}

\def \tr {\textrm{tr}}

\def \and {{~\textrm{and}~}}

\begin{document}

\title{Perturbative distinguishability of black hole microstates from\\AdS/CFT correspondence}

%\begin{CJK*}{GB}{gbsn}
%\begin{CJK*}{UTF8}{gbsn}

\author{Jiaju Zhang}
\email{jiajuzhang@tju.edu.cn}
\affiliation{Center for Joint Quantum Studies and Department of Physics, School of Science,
Tianjin University, 135 Yaguan Road, Tianjin 300350, China}

\begin{abstract}

  We establish direct evidence for the perturbative distinguishability between black hole microstates and thermal states using the AdS/CFT correspondence. In two-dimensional holographic conformal field theories, we obtain the subsystem fidelity and quantum Jensen-Shannon divergence, both of which provide rigorous lower and upper bounds for subsystem trace distance. This result demonstrates that perturbative quantum gravity corrections break semiclassical indistinguishability, thereby supporting the recovery of information even from a small amount of the Hawking radiation.

\end{abstract}

\maketitle

\section{Introduction}

The black hole information paradox \cite{Hawking:1974sw,Hawking:1976ra}, the apparent conflict between unitary quantum evolution and thermal Hawking radiation, remains a cornerstone challenge in quantum gravity. While the AdS/CFT correspondence \cite{Maldacena:1997re,Witten:1998qj,Gubser:1998bc,Aharony:1999ti} provides a holographic framework for unitary evolution, translating this principle into operational distinguishability of microstates and thermal states requires confronting deep questions about semiclassical observables. Holographic entanglement entropy \cite{Ryu:2006bv,Hubeny:2007xt,Faulkner:2013ana} and recent advances of the island formula \cite{Penington:2019npb,Almheiri:2019psf} have resolved the entropy paradox via the Page curve \cite{Page:1993df}, yet fundamental gaps persist \cite{Kudler-Flam:2021rpr,Kudler-Flam:2021alo}: How do quantum gravity corrections encode microscopic state differences into subsystems accessible to observers?

Within AdS$_3$/CFT$_2$, black hole microstates are dual to high-energy eigenstates in the boundary conformal field theory (CFT) \cite{Strominger:1997eq}, while thermal ensembles describe equilibrium black holes with Bekenstein-Hawking entropy. Semiclassical arguments suggest local indistinguishability between these states, a manifestation of the eigenstate thermalization hypothesis (ETH) \cite{Deutsch:1991msp,Srednicki:1994mfb,Deutsch:2018thx,Garrison:2015lva,Dymarsky:2016ntg}, where few-body observables in chaotic systems lose memory of microscopic details. This thermalization persists at leading order in holographic CFTs: highly excited primary states exhibit entanglement entropy matching thermal predictions at $O(c)$ \cite{Fitzpatrick:2014vua,Fitzpatrick:2015zha}, with the large central charge $c = \frac{3R}{2G}$ \cite{Brown:1986nw} encoding the AdS$_3$ radius $R$ and Newton constant $G$. The distinction becomes critical for small subsystems: while the Holevo information \cite{Bao:2017guc} suggests classical indistinguishability at leading order, quantum corrections may encode state-specific signatures through $1/c$ corrections. Results of R\'enyi entropy \cite{Lashkari:2016vgj,Lin:2016dxa}, entanglement entropy and subsystem relative entropy \cite{He:2017vyf}, subsystem Schatten distances \cite{Basu:2017kzo,He:2017txy}, and Holevo information \cite{Guo:2018djz} indicate quantum gravity distinctions that are invisible to semiclassical probes.

Fixed area states \cite{Akers:2018fow,Dong:2018seb,Dong:2019piw}, bulk geometries with constrained extremal surface areas, offer insights into state distinguishability. Studies reveal nonperturbative distinguishability scaling \cite{Kudler-Flam:2021rpr,Kudler-Flam:2021alo}, but status of the fixed area states as superpositions of eigenstates introduces ambiguities dependent on microscopic coefficients. In contrast, primary states in CFT are exact energy eigenstates, providing a natural arena to test intrinsic microstate properties.

The subsystem distinguishability of black hole microstates faces three unresolved challenges. Prior analyses focused on entropy differences \cite{Lashkari:2016vgj,Lin:2016dxa,He:2017vyf}, which lack direct interpretation as distinguishability measures, or Schatten norms \cite{Basu:2017kzo,He:2017txy}, which are not effective measures of distinguishability in high-dimensional Hilbert spaces \cite{Zhang:2020mjv}. Furthermore, fixed area state results depend on superposition details, obscuring universal quantum gravitational effects. Moreover, nonperturbative effects cannot explain observed $1/c$ corrections in holographic CFTs (one can see for example the footnote 17 in \cite{Kudler-Flam:2021alo}).

To resolve these issues, in this letter we compute directly the subsystem fidelity \cite{Nielsen:2010oan} and quantum Jensen-Shannon divergence (QJSD) \cite{Majtey:2005ne,Virosztek:2021fci} between primary and thermal states in 2D holographic CFTs.
Our methodology integrates three synergistic theoretical tools. Firstly, short-interval expansion leverages twist operators \cite{Calabrese:2004eu,Cardy:2007mb,Calabrese:2009qy} and their operator product expansion (OPE) \cite{Headrick:2010zt,Calabrese:2010he,Rajabpour:2011pt,Chen:2013kpa,Lin:2016dxa,Chen:2016lbu,Ruggiero:2018hyl} in two-dimensional conformal field theory to derive closed-form analytic expressions for subsystem fidelity and QJSD, establishing their quantitative roles in microstate distinguishability. Secondly, various inequalities provide upper and lower bounds for trace distance $D(\rho_A, \sigma_A)$ through fidelity $F(\rho_A, \sigma_A)$ and QJSD $J(\rho_A, \sigma_A)$, bypassing computationally intractable trace norm calculations while preserving operational interpretability in high-dimensional Hilbert spaces. Thirdly, large $c$ expansions isolate $1/c$ corrections by prioritizing vacuum module dominance, thereby disentangling quantum gravitational signatures from universal classical parts in holographic systems.

For primary states $|\phi\rangle$, $|\psi\rangle$ with conformal weight difference $h_\phi-h_\psi=O(c^0)$ in a two-dimensional (2D) holographic CFT, we obtain the bound for the subsystem trace distance
\be
O(\ell^4/c) \lesssim D(\r_{A,\phi},\r_{A,\psi}) \lesssim O(\ell^2/c^{1/2}),
\ee
where $\ell$ is the size of the subsystem $A$. The large $c$ scaling breaks semiclassical indistinguishability at perturbative $1/c$ level.
For a primary state $|\phi\rangle$ and thermal state $\rho_\beta$ with matching energy, we get
\be
O(c^0\ell^8) \lesssim D(\r_{A,\phi},\r_{A,\b}) \lesssim O(c^0\ell^4).
\ee
demonstrating quantum distinguishability despite identical classical geometries.
Physically, our results imply that quantum gravity encodes sufficient information in small Hawking radiation subsystems to distinguish microstates. These findings directly challenge the eigenstate thermalization hypothesis (ETH) for holographic systems \cite{Deutsch:2018thx,Dymarsky:2018lhf}, along with previous evidence in the literature \cite{Lashkari:2016vgj,Lin:2016dxa,He:2017vyf,Basu:2017kzo,He:2017txy,Guo:2018djz}, necessitating generalized Gibbs ensembles (GGEs) \cite{Rigol:2006jrd,Cardy:2015xaa} with Korteweg-de Vries (KdV) charges \cite{Bazhanov:1994ft,Cardy:2015xaa,Maloney:2018hdg}, a framework supported by identity conformal family operators \cite{Dymarsky:2019etq} and large subsystem R'enyi entropy \cite{Chen:2024lji,Chen:2024ysb}.

\section{Quantum state distinguishability measures}

The trace distance between two quantum states with density matrices $\r$ and $\s$ is \cite{Nielsen:2010oan}
\be
D(\r,\s)=\f12 \tr|\r-\s|.
\ee
By definition, $0 \leq D(\r,\s) \leq 1$, where $D=0$ implies identical states and $D=1$ implies orthogonal states. The trace distance quantifies the distinguishability of two states under optimal measurements. Although the trace distance possesses a clear operational meaning, it is difficult to compute for large-dimensional Hilbert spaces.

The fidelity between $\r$ and $\s$ is defined as \cite{Nielsen:2010oan}
\be
F(\r,\s)= \tr\sqrt{\sqrt{\r}\s\sqrt{\r}}.
\ee
By definition, $0 \leq F(\r,\s) \leq 1$, where $F=1$ implies identical states and $F=0$ implies orthogonal states. Fidelity measures the closeness of two quantum states.

The trace distance and fidelity are equivalent metrics, linked by the Fuchs-van de Graaf inequalities \cite{Nielsen:2010oan}
\be \label{theinequalities1}
1-F(\r,\s) \leq D(\r,\s) \leq \sqrt{1-F(\r,\s)^2}.
\ee
These inequalities show that trace distance and fidelity provide complementary bounds on the distinguishability of quantum states.

The relative entropy (Kullback-Leibler divergence) between $\r$ and $\s$ is
\be
S(\r,\s)=\tr(\r\log\r)-\tr(\r\log\s).
\ee
By definition, $S(\r,\s)\geq0$, and it is only well-defined when the support of $\r$ is contained within the support of $\s$ (otherwise, $S(\r,\s)=+\inf$).
%Relative entropy quantifies the information loss when approximating ρ with σ.

The Quantum Jensen-Shannon Divergence (QJSD) \cite{Majtey:2005ne} is a symmetric, bounded and smooth measure defined as
\be
J(\r,\s) = S\Big(\f{\r+\s}{2}\Big)-\f12 S(\r)-\f12 S(\s),
\ee
where $S(\r)=-\tr(\r\log\r)$ is the von Neumann entropy. The QJSD can also be written in terms of the relative entropy as
\be
J(\r,\s) = \f12 \Big[ S\Big(\r\Big\|\f{\r+\s}{2}\Big) + S\Big(\s\Big\|\f{\r+\s}{2}\Big) \Big].
\ee
By definition, $0 \leq J(\r,\s) \leq \log 2$. The square root of QJSD is a rigorous metric for general mixed quantum states \cite{Lamberti:2008ptb,Briet:2008lcr,Sra:2019twg,Virosztek:2021fci}.

The QJSD and trace distance are also equivalent metrics, linked by the following inequalities
\be \label{theinequalities2}
\f{J(\r,\s)}{\log 2} \leq D(\r,\s) \leq \sqrt{2 J(\r,\s)}.
\ee
These bounds arise from Pinker's inequality \cite{Watrous:2018rgz} $D(\r,\s) \leq \sqrt{\f12 S(\r\|\s)}$ and a result in \cite{Briet:2008lcr} $J(\r,\s)\leq (\log 2) D(\r,\s)$.

\section{Subsystem distinguishability in 2D holographic CFTs}

Consider a subsystem $A$ of length $\ell$ in a 2D holographic CFT with total spatial length $L$. Let $\r$ and $\s$ be two quantum states, with corresponding reduced density matrices (RDMs) $\r_A$ and $\s_A$ obtained by tracing out the complement of subsystem $A$.

By employing twist operators \cite{Calabrese:2004eu,Cardy:2007mb,Calabrese:2009qy} and their operator product expansion (OPE) \cite{Headrick:2010zt,Calabrese:2010he,Rajabpour:2011pt,Chen:2013kpa,Ruggiero:2018hyl,Lin:2016dxa,Chen:2016lbu}, we derive the short interval expansion of the fidelity \cite{Sui:2025qve}
\bea \label{FrAsA}
&& F(\r_A,\s_A) = 1- \sum_\cX \Big[
                  \f{(2\D_\mX)!\ell^{2\D_\cX}}{2^{4\D_\mX+3}(\D_\mX!)^2}
                  \f{( \lag\cX\rag_\r - \lag\cX\rag_\s )^2}{\ii^{2s_\cX}\a_\cX} \nn\\
&& \phantom{F(\r_A,\s_A)=}
              +o(\ell^{2\D_\cX}) \Big],
\eea
where the sum runs over all quasiprimary operators $\cX$ characterized by conformal weights $(h_\cX,\bar h_\cX)$, scaling dimension $\D_\cX=h_\cX+\bar h_\cX$ and spin $s_\cX=h_\cX-\bar h_\cX$.
The coefficient $\a_\cX$ is related to the normalization of the operator $\cX$, and on the complex plane $\mathbb{C}$ the two-point function is given by
\be
\lag \cX(z_1,\bar z_1) \cX(z_2,\bar z_2) \rag_{\mathbb{C}} = \f{\a_\cX}{(z_1-z_2)^{2h_\cX}(\bar z_1-\bar z_2)^{2\bar h_\cX}}.
\ee
The fidelity (\ref{FrAsA}) is consistent with the analytical and numerical results in free boson and fermion theories, XX chain and critical Ising chain \cite{Zhang:2019wqo,Zhang:2019itb,Zhang:2022nuh}.
Furthermore, in \cite{He:2017txy} the subsystem QJSD was derived as
\bea \label{JrAsA}
&& J(\r_A,\s_A) = \sum_\cX \Big[
                  \f{\sqrt{\pi}\G(\D_\mX+1)\ell^{2\D_X}}{2^{2(\D_\mX+2)}\G(\D_\mX+\f32)}
                  \f{( \lag\cX\rag_\r - \lag\cX\rag_\s )^2}{\ii^{2s_\cX}\a_\cX} \nn\\
&& \phantom{J(\r_A,\s_A)=}
              +o(\ell^{2\D_\cX}) \Big],
\eea
with $\G(x)$ denoting the Gamma function.

We choose $\cX$ being one of the quasiprimary operators $\cX$ satisfying $\lag\cX\rag_\r \neq \lag\cX\rag_\s$ with the smallest scaling dimension $\D_\cX$, and suppose that at the scaling dimension $\D_\cX$ there is no other quasiprimary operators $\cY$ satisfying $\lag\cY\rag_\r \neq \lag\cY\rag_\s$.
Under such condition, the subsystem trace distance is \cite{Zhang:2019wqo,Zhang:2019itb}
\be \label{DrAsA}
D(\r_A,\s_A) = \f{x_\cX\ell^{\D_\cX}}{2} \Big| \f{\lag\cX\rag_\r - \lag\cX\rag_\s}{\sr{\a_\cX}} \Big|
               + o(\ell^{\D_\cX}),
\ee
with the coefficient
\be \label{zX}
x_\cX = \lim_{p \to \f12} \f{\ii^{2 p s_\cX}}{\a_\cX^{p}}
                          \Big\lag \prod_{j=0}^{2p-1} \big[ f_j^{h_\cX} \bar f_j^{\bar h_\cX} \cX(f_j,\bar f_j) \big] \Big\rag_{\mathbb{C}}, ~
f_j \equiv \ep^{\f{\pi\ii j}{p}}.
\ee
In cases where the scaling dimension $\D_\cX$ is degenerate, the form of the trace distance needs to be adjusted, although the scaling behavior $\ell^{\D_\cX}$ remains unchanged.

The coefficient $x_\cX$ is generally difficult to calculate.
However, there are universal upper bounds of of the coefficient $z_\cX$ from inequalities (\ref{theinequalities1}) and (\ref{theinequalities2})
\be \label{constraint1}
x_\cX \leq \frac{\sqrt{(2 \Delta )!}}{2^{2\Delta}\Delta!},
\ee
\be \label{constraint2}
x_\cX \leq \sqrt{\frac{\sqrt{\pi} \G (\Delta +1)}{ 2^{2 \Delta +1} \G (\Delta +\frac{3}{2})}},
\ee
with $\G(x)$ denoting the Gamma function.
The constraint (\ref{constraint1}) from the fidelity is tighter than the constraint (\ref{constraint2}) from the QJSD, while the latter is the same as the constraint in \cite{Zhang:2019itb} obtained from the relative entropy.
Besides, in large $c$ limit, we can show that
\be \label{xTxcA}
x_T \sim O(c^0), ~ x_\cA \sim O(c^0),
\ee
where $T$ is the holomorphic stress tensor and $\cA=(TT)-\f{3}{10}\p^2T$ is a quasiprimay operator at level 4.

The expressions of the subsystem fidelity (\ref{FrAsA}), QJSD (\ref{JrAsA}) and trace distance (\ref{DrAsA}) provide a systematic framework for quantifying the differences between the RDMs $\r_A$ and $\s_A$ of subsystem $A$, capturing the distinguishability of black hole microstates.

\section{Two black hole microstates}

The 2D holographic CFT with a large central charge $c=\f{3R}{2G}$ \cite{Brown:1986nw} is dual to AdS$_3$ quantum gravity, characterized by Newton's constant $G$ and the AdS radius $R$. In this context, a primary state $|\phi\rag$ with large conformal weights $(h_\phi,\bar h_\phi)$, where $h_\phi=\bar h_\phi=c \e_\phi$ and $\e_\phi>\f{1}{24}$, corresponds to a microstate of a non-rotating Ba\~nados-Teitelboim-Zanelli (BTZ) black hole \cite{Banados:1992wn}. Given that the holomorphic and anti-holomorphic sectors decouple, the subsequent analysis will focus solely on the contributions from the holomorphic sector.

We consider two primary states $|\phi\rag$ and $|\psi\rag$ with $h_{\phi}=c \e_{\phi}$ and $h_{\psi}=c \e_{\psi}$, where $\e_{\phi}\neq\e_{\psi}$. The contributions from the holomorphic vacuum conformal family to the subsystem fidelity and QJSD are
\be
F(\r_{A,\phi},\r_{A,\psi})=1-\frac{3 \pi ^4 c \ell ^4 (\e_{\phi}-\e_{\psi})^2}{32 L^4}+ o(\ell^4),
\ee
\be
J(\r_{A,\phi},\r_{A,\psi})=\frac{2 \pi ^4 c \ell ^4 (\e_{\phi}-\e_{\psi})^2}{15 L^4} + o(\ell^4).
\ee
For finite values of $\e_{\phi}$, $\e_{\psi}$ and $\e_{\phi}-\e_{\psi}\sim O(c^0)$ in large $c$ limit, we have
\be
1-F(\r_{A,\phi},\r_{A,\psi}) \sim O(c\ell^4),
\ee
\be
J(\r_{A,\phi},\r_{A,\psi}) \sim O(c\ell^4).
\ee
According to the inequalities (\ref{theinequalities1}) and (\ref{theinequalities2}), the subsystem trace distance is bounded as
\be \label{bounds}
O(c\ell^4) \lesssim D(\r_{A,\phi},\r_{A,\psi}) \lesssim O(c^{1/2}\ell^2).
\ee
This result is intuitive, as two black holes with different masses have distinct metrics and are classically distinguishable. Although it is challenging to evaluate $D(\r_{A,\phi},\r_{A,\psi})$ explicitly, given (\ref{xTxcA}) the expression for the subsystem trace distance (\ref{DrAsA}) suggests
\be
D(\r_{A,\phi},\r_{A,\psi}) = O(c^{1/2}\ell^2),
\ee
which is consistent with the bounds (\ref{bounds}).

We then consider two primary states $|\phi\rag$ and $|\psi\rag$ with equating leading order conformal weights $h_{\phi}=c \e_{\phi}+\d_{\phi}$ and $h_{\psi}=c \e_{\phi}+\d_{\psi}$.
For finite values of $\d_{\phi}\sim O(c^0)$, $\d_{\psi}\sim O(c^0)$ and $\d_{\phi}-\d_{\psi}\sim O(c^0)$ in the large $c$ limit, the subsystem fidelity and QSJD are given by
\be
F(\r_{A,\phi},\r_{A,\psi}) = 1 - \frac{3 \pi ^4 \ell ^4 (\d_{\phi}-\d_{\psi})^2}{32 c L^4} + o(\ell^4),
\ee
\be
J(\r_{A,\phi},\r_{A,\psi}) = \frac{2 \pi ^4 \ell ^4 (\d_{\phi}-\d_{\psi})^2}{15 c L^4} + o(\ell^4).
\ee
Accordingly, the subsystem trace distance is bounded as
\be
O(\ell^4/c) \lesssim D(\r_{A,\phi},\r_{A,\psi}) \lesssim O(\ell^2/c^{1/2}).
\ee
The formula (\ref{DrAsA}) indicates that
\be
D(\r_{A,\phi},\r_{A,\psi}) = O(\ell^2/c^{1/2}).
\ee
In this case, the two black hole microstates are classically indistinguishable, but they can be distinguished through perturbative quantum corrections. This result highlights the subtle quantum effects that allow for the differentiation of black hole microstates, even when classical metrics are identical.

For two primary states $|\phi\rag$ and $|\psi\rag$ with exactly the same conformal weights $h_\phi=h_\psi$ and $\bar h_\phi=\bar h_\psi$, the vacuum conformal family does not contribute to the short interval expansion of the subsystem fidelity and QJSD. With $\phi$ and $\psi$ being hermitian and the normalization $\a_\phi=\a_\psi=1$, we obtain
\bea
&& \hspace{-5mm} F(\r_{A,\phi},\r_{A,\psi}) = 1- \sum_\cX
                                \Big[ \f{(2\D_\mX)!}{2^{2\D_\mX+3}(\D_\mX!)^2}
                                      \Big( \f{\pi\ell}{L} \Big)^{2\D_\cX} \nn\\
&& \hspace{-5mm} \phantom{F(\r_{A,\phi},\r_{A,\psi})=} \times
                                      \f{( C_{\phi\cX\phi}-C_{\psi\cX\psi} )^2}{\a_\cX}
                                     +o(\ell^{2\D_\cX})
                                \Big],
\eea
\bea
&& \hspace{-5mm} J(\r_{A,\phi},\r_{A,\psi}) = \sum_\cX
                                \Big[ \f{\sqrt{\pi}\G(\D_\mX+1)}{16\G(\D_\mX+\f32)}
                                      \Big( \f{\pi\ell}{L} \Big)^{2\D_\cX} \nn\\
&& \hspace{-5mm} \phantom{J(\r_{A,\phi},\r_{A,\psi})=} \times
                                      \f{( C_{\phi\cX\phi}-C_{\psi\cX\psi} )^2}{\a_\cX}
                                     +o(\ell^{2\D_\cX})
                                \Big],
\eea
with $C_{\phi\cX\phi}$ and $C_{\psi\cX\psi}$ being the structure constants. Note that $1-F(\r_{A,\phi},\r_{A,\psi})$ and $J(\r_{A,\phi},\r_{A,\psi})$ are at the same leading order, whose behavior in the large $c$ limit depends on the details of the structure constants in the 2D CFT.

\section{Black hole microstate and thermal state}

Finally, we examine a primary state $|\phi\rag$ with conformal weight $h_\phi=c\e_\phi$ and a thermal state $\r_\b=\ep^{-\b H}/Z(\b)$ with $Z(\b)=\tr(\ep^{-\b H})$. According to \cite{Fitzpatrick:2014vua,Fitzpatrick:2015zha}, when
\be
\b=\f{L}{\sqrt{24\e_\phi-1}},
\ee
the two states have the same energy and the same order $O(c)$ entanglement entropy. However, \cite{He:2017vyf} shows that with $1/c$ corrections, their entanglement entropies differ, which also indicates that the subsystem relative entropy $S(\r_{A,\phi}\|\r_{A,\b})$ scales as $O(c^0\ell^8)$.

We obtain the subsystem fidelity as
\be
F(\r_{A,\phi},\r_{A,\b}) = 1- \frac{7 \pi ^8 c \ell ^8 (22 \epsilon_\phi-1)^2 \epsilon_\phi ^2}{512 (5 c+22) L^8} + o(\ell^8).
\ee
The subsystem QJSD was derived in \cite{He:2017txy} as
\be
J(\r_{A,\phi},\r_{A,\b}) = \frac{32 \pi^8 c \ell^8 \epsilon_{\phi}^2 (22 \epsilon_{\phi}-1)^2}
                                    {1575 (5 c+22) L^8} + o(\ell^8).
\ee
Applying the inequalities (\ref{theinequalities1}) and (\ref{theinequalities2}), we obtain the subsystem trace distance bounded by
\be
O(c^0\ell^8) \lesssim D(\r_{A,\phi},\r_{A,\b}) \lesssim O(c^0\ell^4).
\ee
The formula for the subsystem trace distance (\ref{DrAsA}) indicates
\be
D(\r_{A,\phi},\r_{A,\b}) = O(c^0\ell^4).
\ee
This perturbative distinguishability demonstrates that black hole microstates encode more information than a thermal state. Although they are classically indistinguishable, quantum effects break this degeneracy, allowing information to be preserved.\\

\section{Discussions and Conclusions}

Using subsystem fidelity and QJSD as distinguishability metrics on exact primary states and thermal states, we established universal $1/c$ scaling of subsystem trace distance, proving state distinguishability persists at large central charge for small $\ell/L$. Our results from holographic CFTs establish that quantum gravitational corrections can induce perturbative distinguishability between black hole microstates and thermal states, even for small subsystems, thus solving the puzzle mentioned in the footnote 17 of \cite{Kudler-Flam:2021alo}.

The perturbative distinguishability of microstates via small subsystems resolves a key obstacle to information retrieval: while semiclassical arguments forbid distinguishing microstates from Hawking radiation, quantum corrections at $O(c^0)$ order encode sufficient information in finite-dimensional subsystems. This mechanism complements the island formula by providing an operational pathway to reconstruct microstate details from radiation.

The failure of traditional ETH for primary states, evident from various quantities in the literature \cite{Lashkari:2016vgj,Lin:2016dxa,He:2017vyf,Basu:2017kzo,He:2017txy,Guo:2018djz} and the subsystem trace distances between microstates and thermal ensembles in this work and \cite{Sui:2025qve}, necessitates extending thermalization to generalized Gibbs ensembles (GGEs) incorporating KdV charges. This aligns with recent findings in large-$c$ CFTs \cite{Dymarsky:2019etq,Chen:2024lji,Chen:2024ysb}, where KdV conservation laws govern subsystem thermalization.

While primary states dominate the microcanonical ensemble, descendants \cite{Guo:2018pvi,Guo:2018fnv,Brehm:2020zri} may exhibit enhanced distinguishability due to Virasoro hair. Extending our framework to include descendant contributions could reveal new distinguishability structures. While our analysis focuses on 2D holographic CFTs, analogous results for higher-dimensional CFTs could test the universality of perturbative distinguishability of black holes.

In summary, this work demonstrates that quantum gravity corrections in holographic CFTs fundamentally alter the semiclassical notion of black hole thermalization. By establishing perturbative distinguishability as a generic feature of AdS$_3$/CFT$_2$, we bridge the gap between information-theoretic principles and geometric unitarity, offering a concrete step toward fully resolving the black hole information paradox.

\section*{Acknowledgements}

%{\it Acknowledgements.}
%The author thanks ... for reading a previous versions of the draft and helpful comments and discussions.
The author thanks Li-Ming Cao, Bin Chen, Anatoly Dymarsky, Wu-zhong Guo, Song He, Yan Liu, Jian-Xin Lu, Jia-Rui Sun, Shao-Jiang Wang, Jie-qiang Wu, and Li-Xin Xu for their helpful discussions and comments. The author acknowledges support from the National Natural Science Foundation of China (NSFC) under Grant No.~12205217, and from the Tianjin University Self-Innovation Fund Extreme Basic Research Project under Grant No.~2025XJ21-0007.

\providecommand{\href}[2]{#2}\begingroup\raggedright\endgroup

%\bibliographystyle{D:/00.bibx/JHEPx}
%\bibliography{D:/00.bibx/2025,D:/00.bibx/2024,D:/00.bibx/2023,D:/00.bibx/2022,D:/00.bibx/2021,D:/00.bibx/2020,D:/00.bibx/2019,D:/00.bibx/2018,D:/00.bibx/1960,D:/00.bibx/1970,D:/00.bibx/1980,D:/00.bibx/1990,D:/00.bibx/1995,D:/00.bibx/1996,D:/00.bibx/1997,D:/00.bibx/1998,D:/00.bibx/1999,D:/00.bibx/2000,D:/00.bibx/2001,D:/00.bibx/2002,D:/00.bibx/2003,D:/00.bibx/2004,D:/00.bibx/2005,D:/00.bibx/2006,D:/00.bibx/2007,D:/00.bibx/2008,D:/00.bibx/2009,D:/00.bibx/2010,D:/00.bibx/2011,D:/00.bibx/2012,D:/00.bibx/2013,D:/00.bibx/2014,D:/00.bibx/2015,D:/00.bibx/2016,D:/00.bibx/2017,D:/00.bibx/book,D:/00.bibx/work,D:/00.bibx/thesis}

%\clearpage

%\newpage

%\onecolumngrid

%change the format of page, section, equation, figure and table%
%\renewcommand\thepage{S\arabic{page}}
%\renewcommand\thesection{S\arabic{section}}
%\renewcommand{\thesubsection}{S\arabic{subsection}}
%\renewcommand\theequation{S\arabic{equation}}
%\renewcommand\thefigure{S\arabic{figure}}
%\renewcommand\thetable{S\arabic{table}}

%\setcounter{page}{1}
%\setcounter{section}{0}
%\setcounter{equation}{0}
%\setcounter{figure}{0}
%\setcounter{table}{0}

%\begin{appendices}
%\addtocontents{toc}{\protect\setcounter{tocdepth}{1}}
%\renewcommand{\theequation}{\thesection.\arabic{equation}}

%\begin{CJK*}{GB}{gbsn}
%\begin{CJK*}{UTF8}{gbsn}

%\begin{center}
%{\large \textbf{Supplemental Materials: {Universal Volume-law Entanglement Fragmentation of Quasiparticles}}}% \\~\\
%Jiaju Zhang %(张甲举)
%\end{center}

%\end{CJK*}

\end{document}